\documentclass[runningheads]{llncs}
\usepackage{cite}
 \usepackage{color}
 \usepackage{graphicx}
 \usepackage{array}
\begin{document}
\title{AstroServ: Distributed Database for Serving Large-Scale Full Life-Cycle Astronomical Data}
\titlerunning{AstroServ: Serving Large-Scale Full Life-Cycle Astronomical Data}
%
\author{Chen Yang\inst{1} \and Xiaofeng Meng \inst{1}\thanks{Corresponding author is Xiaofeng Meng} \and Zhihui Du\inst{2} \and JiaMing Qiu\inst{2} \and Kenan Liang\inst{3} \and Yongjie Du\inst{1} \and Zhiqiang Duan\inst{1} \and Xiaobin Ma\inst{2} \and Zhijian Fang \inst{3}}
\authorrunning{Chen Yang et al.}
%
\institute{School of Information, Renmin University, China \and
Department of Computer Science and Technology, Tsinghua University, China \and
School of Computer Science and Technology, Shandong University, China}
\maketitle              
\hyphenation{strict HSR different unit GWAC}
\begin{abstract}
In time-domain astronomy, STLF (Short-Timescale and Large Field-of-view) sky survey is the latest way of sky observation. Compared to traditional sky survey who can only find astronomical phenomena, STLF sky survey can even reveal how short astronomical phenomena evolve. The difference does not only lead the new survey data but also the new analysis style. It requires that database behind STLF sky survey should support continuous analysis on data streaming, real-time analysis on short-term data and complex analysis on long-term historical data. In addition, both insertion and query latencies have strict requirements to ensure that scientific phenomena can be discovered. However, the existing databases cannot support our scenario. In this paper, we propose AstroServ, a distributed system for analysis and management of large-scale and full life-cycle astronomical data. AstroServ's core components include three data service layers and a query engine. Each data service layer serves for a specific time period of data and query engine can provide the uniform analysis interface on different data. In addition, we also provide many applications including interactive analysis interface and data mining tool to help scientists efficiently use data. The experimental results show that AstroServ can meet the strict performance requirements and the good recognition accuracy.
\end{abstract}
\section{Introduction}
\begin{figure}
  \centering
  \includegraphics[width=5in,height=1.4in]{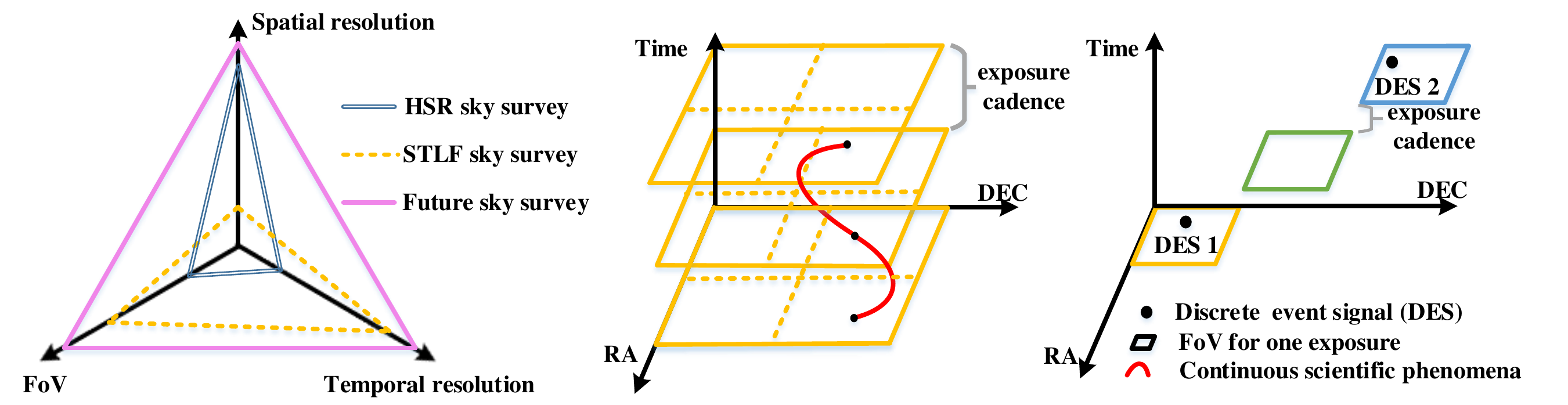}
  \caption{The left is a radar graph, which compares characteristics of different sky surveys. STLF sky survey (middle) can generate
  continuous time-series data of the same star in one night. HSR sky survey (right) can only generate discrete data points in one night. RA and DEC are right ascension and declination, respectively.}\label{fig:GWACDdatavsOthersurvey}
\end{figure}

In recent years, many large optical instruments in time-domain astronomy have brought unprecedented observation capabilities to us.
As shown in Figure \ref{fig:GWACDdatavsOthersurvey}, these instruments have made great progress in three factors, including field-of-view (FoV), spatial resolution and temporal resolution, which means that the telescope can search larger area and more darker objects with higher frequency, respectively. Due to the cost limitation, three factors cannot currently be met at the same time, so that two ways are attempted to design observation instruments, including HSR (High Spatial Resolution) sky survey and STLF sky survey.

The study about HSR sky survey has been around for a long time, such as PTF\cite{nugent2015palomar}, Skymapper\cite{SkyMapper2017}, SDSS\cite{frieman2008the}, Pan-STARRS\cite{Huber2015The}, and LSST \cite{becla2008organizing}. They finish a survey within 3-5 days and scan different regions (about 3-9.6 square degrees) at every exposure. Thus, HSR sky survey cannot catch short and continuous scientific phenomena or only catch discrete scientific events of different observed regions.

STLF sky survey as a new observation approach cannot only simultaneously observe lots of spacial objects, but also a huge amount of data is collected at a high exposure frequency. For example, GWAC\cite{wan2016column} finishes a survey within 15 seconds and scans the same region (about 5000 square degrees) at every exposure. Through STLF sky survey, human being can understand how short astronomical phenomena evolve. Although the new observation ability can help scientists reveal more natural laws, the database systems behind STLF sky survey will face unprecedented challenges.

The new survey data generated by STLF sky survey provides scientists a completely new way to achieve scientific discovery, being very different from HSR sky survey. The databases behind HSR sky survey, such as SkyServer\cite{szalay2002sdss} and Qserv\cite{Wang2011Qserv}, only support the management of long-term historical data. However, continuous time-series data has higher value and supports more analysis methods on them, compared with discrete data points. Thus, scientists often expect to launch an analytical query on streaming data, short-term data and long-term historical data to confirm a scientific phenomenon and issue an alert as soon as possible. It causes the analysis requirement not limited to long-term historical data, but extend to full life-cycle of data.
Obviously, it is more helpful for scientists to reveal scientific laws on site since many astronomical physical phenomena are transient and hard to reproduce, such as microlensing.

It is a challenging work to design a database behind STLF sky survey to support continuous analysis on streaming data, real-time analysis on short-term data and complexity analysis on long-term historical data. We face stringent data challenges as follows.
 \begin{itemize}
 \item \textbf{High density}. Hundreds of thousands of objects will be extracted after an exposure, causing the database to require high throughput. For example, GWAC can simultaneously observe about 3.5 million objects.

  \item \textbf{High frequency}. Many data will be generated in very short time, causing the database to require low request latency.
  For example, GWAC will produce data every 15 seconds so that The ingest latency must be less than 15 seconds and a real-time query must be also finished less than 15 seconds.
  \item \textbf{Tremendous amount}. As time goes on, a lot of data will be accumulated, causing the database to require efficient storage capacity. For example, GWAC can collect 6.7 billion high-dimensional data points per night. Finally, produce about 2.24PB size of data over 10 years. Thus, the storage resource should be used as little as possible from an economic point of view.
\end{itemize}

     \begin{figure}[t]
  \centering
  \includegraphics[width=5in,height=2in]{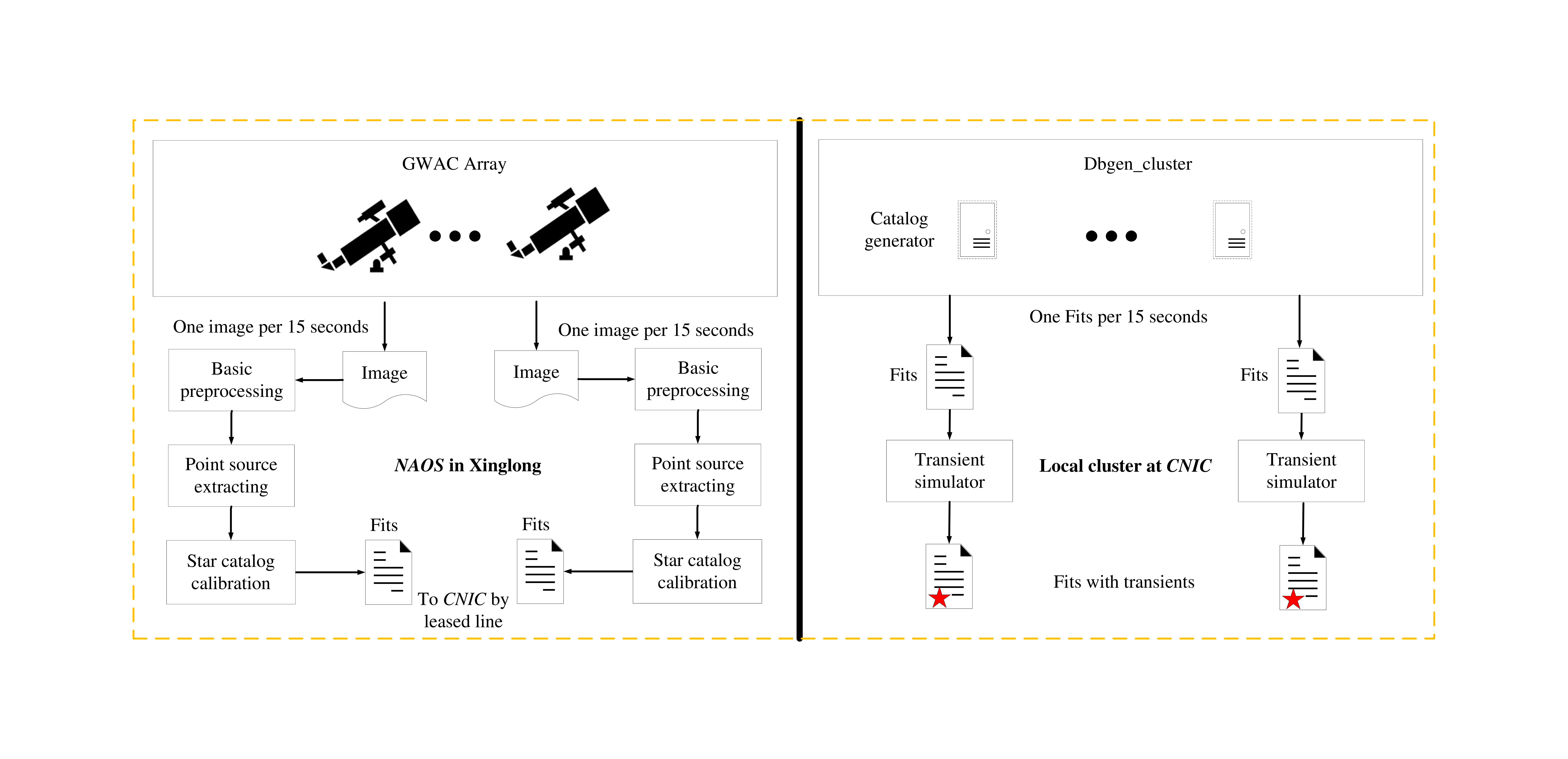}
  \caption{The left is the real data generation pipeline and the right is data generator to simulate GWAC's working mode.}\label{fig:dataSource}
\end{figure}

 The existing astronomical databases are designed for HSR sky survey, so they do not have a good support for continuous analysis and real-time analysis. Other high-performance databases, such as MonetDB\cite{wan2016column} and Hbase\cite{Hbase}, cannot support the scientific analysis. Thus, in this paper we follow GWAC's data feature to design a database AstroServ serving full life-cycle astronomical data of STLF sky survey. AstroServ mainly includes five major contributions as follows.
 \begin{itemize}
   \item \textbf{Detect service on streaming}. It receives the newly arriving data, normalizes data with a scientific pipeline and recognizes abnormal star from huge amounts of objects. This module can follow the status of observation instruments in real time and find early phase of scientific phenomena.
   \item \textbf{Real-time data service}. It receives normalized data and identified abnormal star information and manage one night of data. This module can efficiently ingest them into in-memory store to provide real-time query service by a set of compact data store and index structures.
   \item \textbf{Long-term data service}. It ingests one night of data every time and finally manage all historical data. This module can append data with a high throughput and provide off-line query service by compact data scheme and efficient metadata management.
   \item \textbf{Query engine.} We design a framework to build query engine to ensure that it can query data from both real-time data service and long-term data service with a unified way.
   \item \textbf{Applications.} Combined with core service of our database, we present two applications including interactive analysis interface and off-line integrated detector to help user fast tracking and data mining.
 \end{itemize}
     The rest of the paper is organized as follows. Section \ref{section:background} introduces the GWAC project and data sources. Section \ref{section:Architecture} presents our work. Section \ref{section:experiment} describes our experimental results. Section \ref{section:relatedWork} is related work. Section \ref{section:Summary} summaries our work and presents directions for future work.
\begin{figure}[t]
  \centering
  \includegraphics[width=1.5in,height=2.2in]{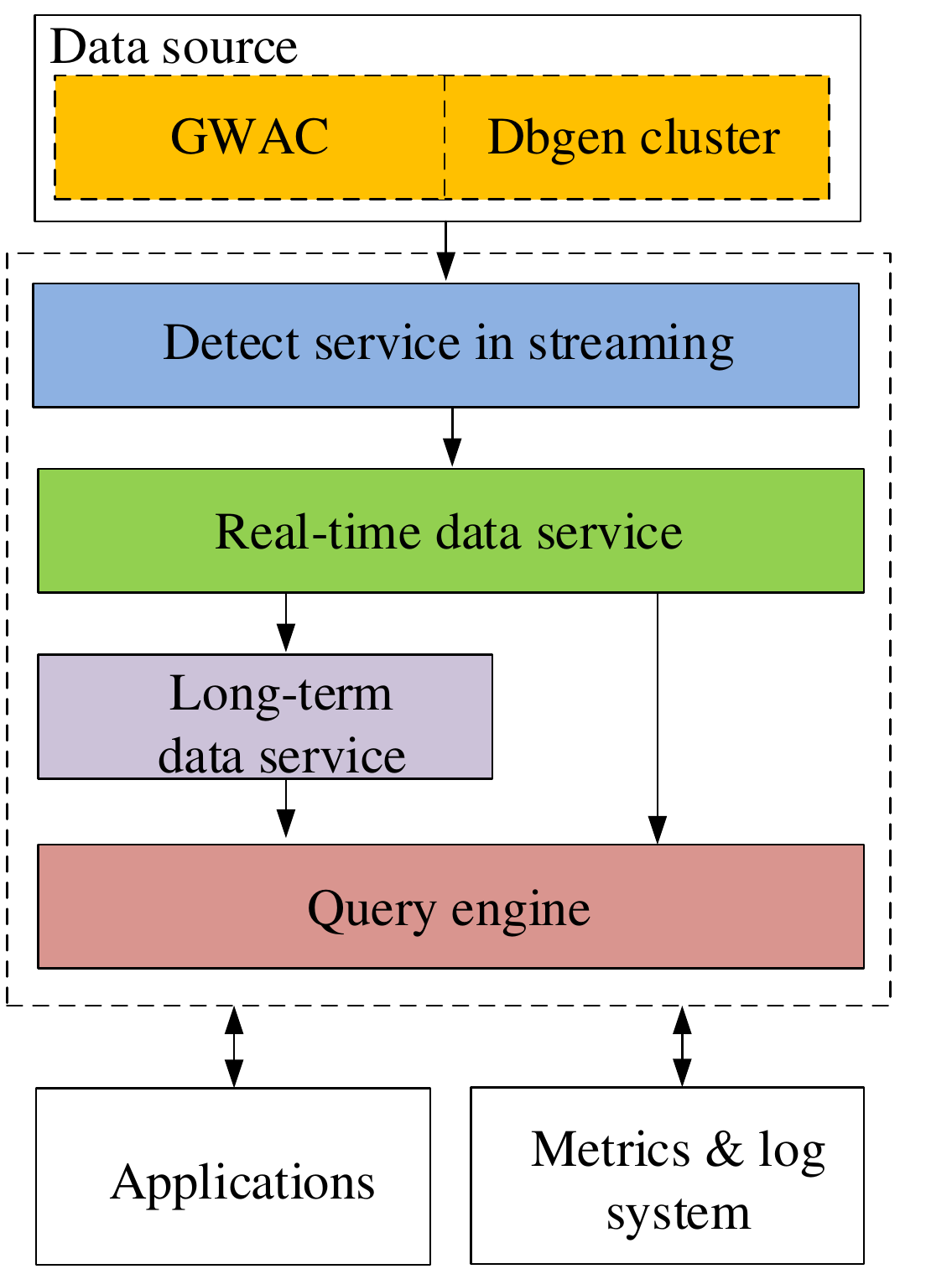}
  \caption{Major components in AstroServ}\label{fig:astroDBArchitecture}
\end{figure}
\section{Background}\label{section:background}
\subsection{GWAC Project}
The Ground-based Wide-Angle Camera array (GWAC) with 20 cameras\footnote[1]{GWAC will eventually expand to 36 cameras.}, which was built in China, was a part of the SVOM space mission. Each camera can observe about 175,600 objects. Thanks to the low exposure cycle and large FoV, the survey cycle of GWAC was equal to the exposure cycle 15 seconds. Thus, it is ideal for searching for optical transients of various types by continuously imaging the same region.

GWAC can produce 3.5 million rows of data per 15 seconds and 6.7 billion rows of data are produced in one night (about 8 hours). 2.24PB size of data will be produced over 10 years. It suggests that the worst performance of our database must meet (1) the throughput is more than 234,133 rows per second in detect service on streaming, (2) the throughput is also more than 234,133 rows per second and the query latency on 6.7 billion rows of data is less than 15 seconds, and (3) long-term data service can also ingest 6.7 billion rows of data into database within 12 hours and can manage 2.24PB data and provide the tolerable query latency.

\subsection{Data Sources}
GWAC as a camera array only produces the image and the image will be transformed into a relational table named as the catalog file through point source extracting, etc, as shown in Figure \ref{fig:dataSource}. The catalog file sent from remote place (e.g., Xinglong) to local servers (e.g., Beijing) will be used as AstroServ's final input. To easily design our database, we first develop a distributed data generator on local cluster where each machine simulates a GWAC's camera and synchronously generates the catalog file.
In addition, we also design a transient simulator to randomly attach the scientific phenomenon to stars (e.g., attaching continuous microlensing to some star) to test our detect service.

\section{AstroServ's Architecture}\label{section:Architecture}

As shown in Figure \ref{fig:astroDBArchitecture}, AstroServ mainly includes six major components, where detect service in streaming, real-time data service, long-term data service and query engine belong to the core components. The core components manage different phases of data and provide the query service. Finally, achieve the full life-cycle management.
\subsection{Detect Service on Streaming}
\begin{figure}[t]
  \centering
  \includegraphics[width=5in,height=1.6in]{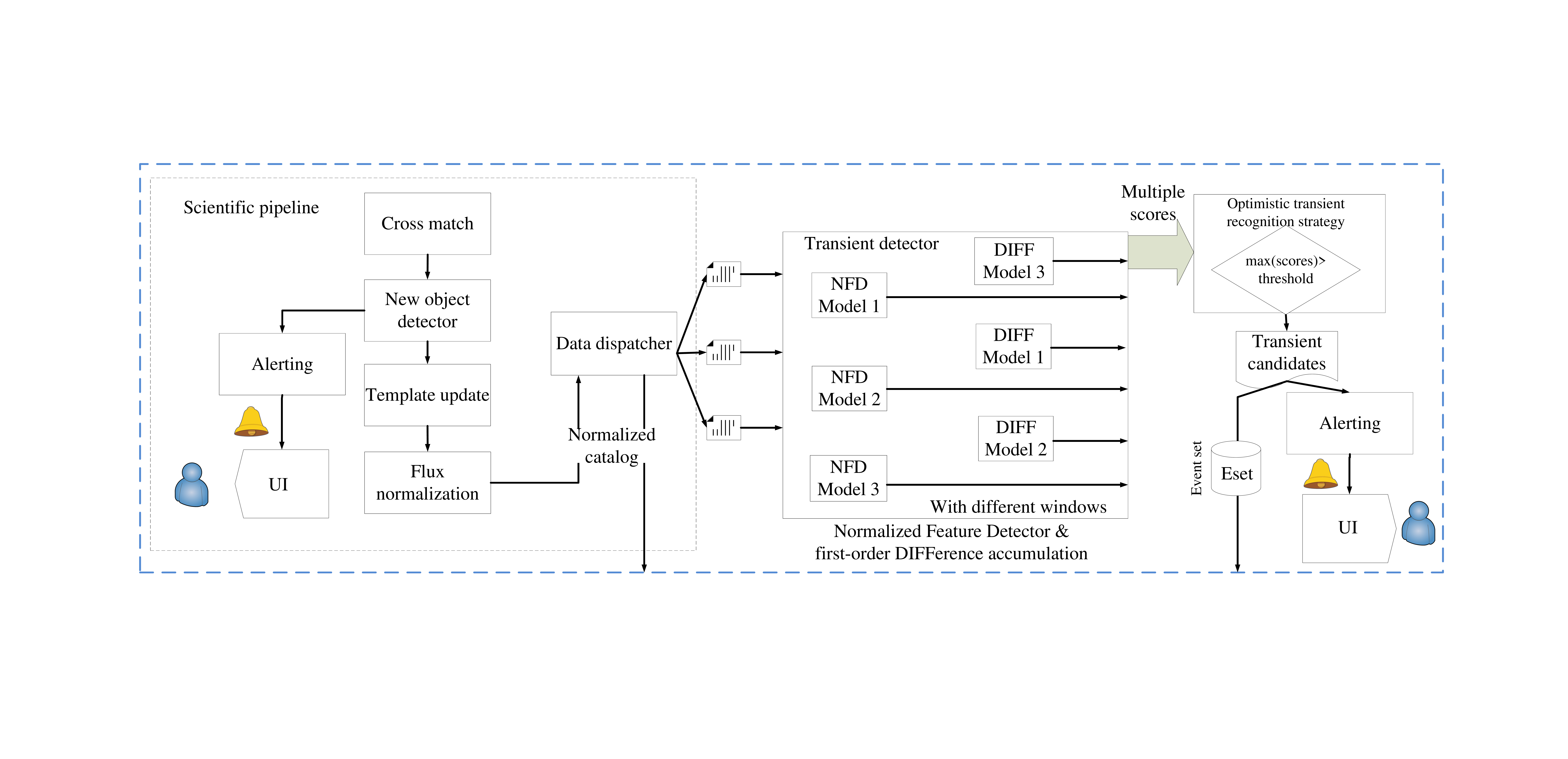}
  \caption{Scientific pipeline and detector (the rest) in detector service on streaming}\label{fig:DetectServiceonStreaming}
\end{figure}

This component follows the ``master-slave" mode and each slave node serves the data generated by a camera, including two functions: (1) scientific pipeline and (2) detector\cite{Feng2017Real}. Scientific pipeline can assign the IDs to stars, find the new stars and normalizes the catalog data to help detector improve the recognition precision. Then, several key steps have to be done. First, the catalog file where stars have no IDs is sent to cross-match module\cite{nieto2007cross}, stars will be given IDs by searching the nearest star in the template table which includes all of identified stars. In cross-match module, we use the pixel coordinates to improve the performance\cite{Xu2013A}. If the star in catalog does not appear in the template, it suggests a new star and AstroServ will send an alerting signal. When the new stars appear, they will be added into the template table. We design an encode strategy to define the star ID and maintain the template table automatically. In addition, flux normalization was carried out through comparison to standard stars in the same fields to correct the magnitude error due to external interference, such as the cloud. After that, the normalized catalog could be just used for transient detection and storage.

The detector is used for recognizing the scientific phenomena. After each exposure, it receives the normalized catalogs and returns the set of abnormal star IDs, called as Eset (Event set). The detector uses the integration framework based on sliding window, including six models to detect different types of scientific phenomena. Six models derive from two methods being NFD (Normalized Feature Detector) and DIFF (first-order DIFFerence accumulation)\cite{nieto2007cross} with different window sizes. Each model will give a score representing the probability of an abnormal star to next module. An optimistic transient recognition strategy will be used to decide Eset according to the maximum score of every stars. Finally, send the alerting signals.

\subsection{Real-Time Data Service}
\begin{figure}[t]
  \centering
  \includegraphics[width=5in,height=1.6in]{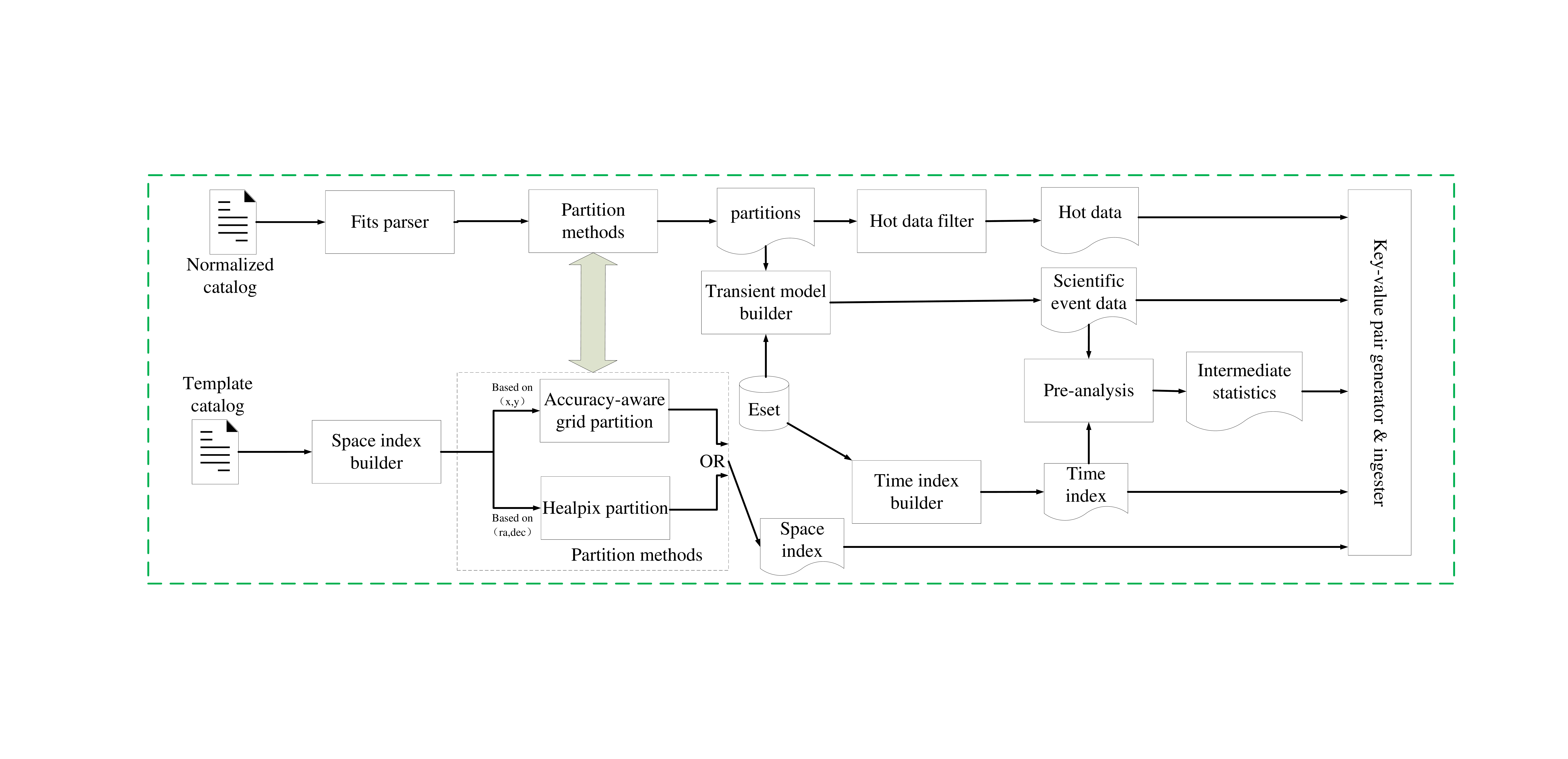}
  \caption{Architecture of real-Time data service}\label{fig:RealtimeDataDervice}
\end{figure}
This component focuses on efficient storage and query of short-term data (e.g., one night of data). For example, one night of data in GWAC has high value, so that the low-latency access to these data can help user analyze scientific phenomena on site. Thus, both the ingest and query latency must be less than the exposure interval. As shown in Figure \ref{fig:RealtimeDataDervice}, we use a set of optimization methods to improve the performance. We build scientific phenomenon model based on key-value schema to organize data to speed up the access to time-series data. In addition, we use partition methods to partition both catalog and template table and physically cluster star data of the same partition to improve the ingest due to less partition keys. Noting that we implement two partition methods: (1) accuracy-aware grid partition presented by us and (2) Healpix partition\cite{Gorski1999The}. Accuracy-aware grid partition supports the approximate query to improve the query performance. For improving the access to scientific phenomenon data, we build a time index based on time-evolving feature. It has the insert-friendly structure and supports the distributed scan. For speeding up the analysis to scientific phenomenon data, we first pre-analyze the abnormal stars inside every partition and generate intermediate statistics to avoid the access to original data. In addition, we also store hot data separately to improve their access. All of generated data are finally transformed into key-value pairs and inserted into a distributed key-value in-memory store. Although we insert multiple types of data but they are clustered into different partitions so that it does not reduce the ingest performance. Using the partition as the insert unit can also reduce the unnecessary data structure overhead.
\subsection{Long-term Data Service}
\begin{figure}[t]
  \centering
  \includegraphics[width=5in,height=2.4in]{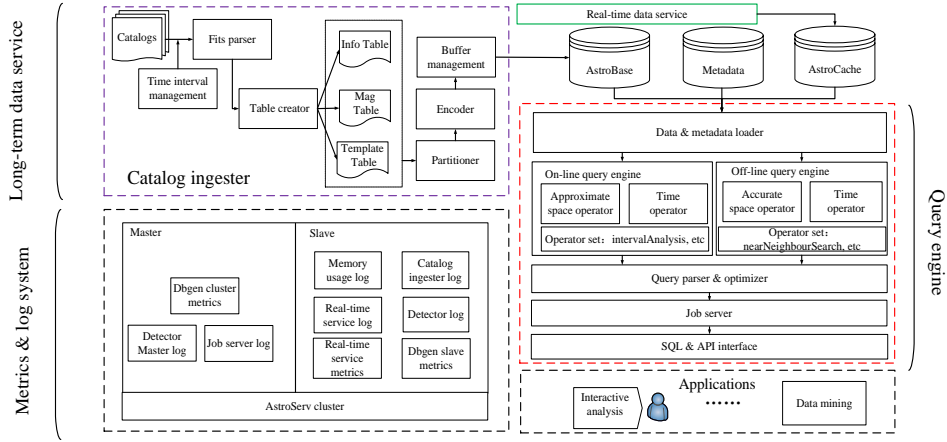}
  \caption{Long-term data service, query engine, metrics \& log system and applications}\label{fig:OtherService}
\end{figure}
This component is actually a catalog ingester which ingests one night of catalogs into a disturbed on-disk store. The major challenge is to ensure that the insert latency is less than 8 hours. Thus, we overlap the processing between real-time data service and long-term data service. As shown in Figure \ref{fig:OtherService}, long-term data service (i.e., time interval management) can dynamically monitor the generation of normalized catalog and ingest catalogs when real-time data service is idle. We totally use three logical tables to keep data: (1) original time-series information except magnitude attribute in Info table, (2) magnitude attribute in Mag table and (3) non-time-series information such as locations in Template table. This organization form can improve the performance of the popular light curve analysis. For saving storage space, we design an encoder to compress data. In addition, the usage of memory buffer can also improve the insert performance.
\subsection{Query Engine}
This component needs to support both the real-time analysis on short-term data and complex analysis on long-term data. Short-term data is kept into AstroCache being a disturbed key-value in-memory store, such as Redis cluster\cite{RedisCluster} and long-term data is kept into AstroBase being a disturbed on-disk store, such as Hbase\cite{Hbase}. Real-time analysis focuses on the scientific phenomena recognized by detector and complex analysis focuses on large-scale star data. We design a query engine integration framework which can
transparently execute queries from different data sources. The framework uses Web service (i.e., Job server module) as the query interface and supports both SQL-like and API to access AstroServ. By query parser, query engine can choose the correct sub-engine to run query. As shown in Figure \ref{fig:OtherService}, our on-line query engine runs approximate query efficiently and can meet the minimum accuracy, but our off-line query engine is accurate. When the query engine is launched, we first load the metadata into memory to improve the query performance and every query job will be run distributedly.
\subsection{Other Components}
\textbf{Metrics \& log system}. AstroServ as a distributed database follows the ``master-slave" mode, so that we monitor services on master and slave nodes, respectively. Different logs are benefit from profiling our system. The major metrics include the size of data in both AstroCache and AstroBase, the Eset's size, the catalog size, the ingest and query latency, etc.

\textbf{Applications}. They mainly obtain data from the core components and help users analyze data in an easier way. Currently, our applications include interactive analysis interface and data mining tool. Interactive analysis interface is a web UI which is ability to graphically display alerting signals and analysis results on an earth model. Data mining tool is a machine learning algorithm integration framework which can recognize long-timescale scientific phenomena from long-term historical data.

\section{Experiments}\label{section:experiment}
\subsection{Experimental Setup}
We evaluate how well AstroServ works by using four metrics: insertion latency, query latency, detector recognition precision and storage usage. We use GWAC\cite{wan2016column} as an example to test our system, in which each observation unit can collect 175,600 objects per 15 sec and an observation experiment lasts 1,920 times (about 8 hours).

We simulate GWAC by using the distributed data generator. It follows the ``master-slave" mode, where each sub-generator simulates an observation unit. A sub-generator produces a catalog file which is a relational table with 175,600 lines and 25 columns per cycle. In our experiment, we assign every sub-generator to run on one machine. In addition, we simulate scientific event signals by setting the Eset size to subject to the geometric distribution and the locations of scientific events to subject to the uniform distribution. The duration of each scientific event is also random. Finally, we simulate 19 observation units on 20 machines (one master), each of which has 12 CPU cores (CPU frequency is low, only 1.6 GHz per core) and 96 GB RAM.

 We also build AstroServ's cluster on the same 20 machines where each slave processes data produced by a sub-generator on the local machine. We use Go and C++ to implement detector service and real-time service, respectively. Redis cluster 3.2.11 as the key-value in-memory store where we launch 120 storage nodes (i.e., master and slave nodes are half of each) and each slave node backs up data of a master node. Hbase 1.3.1 is as the on-disk store. Spark 1.6.3\cite{ApacheSpark} is used for query processing. We will divide the experiment into three parts as follows.
\subsection{Results}
\textbf{Detect service on streaming}. For each slave node, the average detector latency on 17,5600 rows of data is 3.97 seconds and the variance is 1.6 seconds. Noting that the Go's garbage collection may cause the detector latency to be longer. Although it has no much impact on the overall performance, but it needs to be further optimized. We also test the detector's accuracy. We use GWAC's specific parameters to generate 3,240 time-series  as test data and add the scientific phenomena for each time-series. Our's detector accuracy can achieve 73.5\%. In other words, we have ability to detect 2,381 scientific phenomena.

\textbf{Real-time data service}. We ingest 19 catalog files per 15 seconds by real-time data service lasting 1,920 times (8 hours). The average insert latency is 2.35 seconds which is less than our requirement 15 seconds. In addition, our ingest latency is stable and the variance is 0.12 seconds. The size of 8 hour data is 1.15TB, and AstroCache only uses 460GB main memory. It reduces the main memory overhead of our system. As a comparison, Redis cluster without our optimization will consume 2.34TB main memory. The main reason is that (1) we only manage the hot data and scientific phenomenon data instead of the entire data set and (2) we use partition data instead of star data as the storage unit which can cut unnecessary data structures to maintain more keys.

\textbf{Long-term data service}. We prepare one-night of data in advance to test the continuous insertion performance. This component consumes 3.5 hours to ingest these catalogs into AstroBase and about 6 seconds for each catalog. The latency is less than our requirement 8 hours. As a comparison, Hbase without our optimization needs 13 hours to ingest one-night of data and about 24.4 seconds for each catalog. Our optimization can improve 3.7$\times$ insertion performance.

\textbf{Query engine}. For on-line query engine testing, we refer to the LSST telescope discovery ability \cite{becla2008organizing} and decide to simulate the generation of 200,000 scientific phenomena one night. This number is more than it at the real case to test the worst performance of on-line query engine. The worst query latency on 6.7 billion rows of data is 2.72 seconds being less than 15 seconds. For off-line query engine testing, we preliminarily test it on 15.71TB (120 billion rows of data). We assume that the size of result set is less than 1 million rows of data. When the result set is 1.2 million rows of data, the average query latency is 26 seconds.

\section{Related Work}\label{section:relatedWork}
We survey the related work involving (1) astronomical data management system and (2) distributed data store.

\textbf{Astronomical data management system}. In time-domain astronomy, catalogs collected by telescopes had to be stored into a long-term database for complex analysis. SkyServer\cite{szalay2002sdss} for SDSS was built on Microsoft SQL Server. It was primarily responsible for long-term storage (since 2000), complex query and the primary public interface catalogue data from SDSS. SciServer\cite{sciserver} was a major upgrade of SkyServer. SciServer was a collaborative research environment for large-scale data-driven science, but it still worked on long-term data. Qserv\cite{Wang2011Qserv} was a distributed shared-nothing database to manage the LSST catalogs over 10 years. It was designed into a MySQL cluster by a proxy server and a distribution file system xrooted. In addition, PostgrelSQL was used for storing the original data for Skymapper. These databases did not consider the efficient real-time time-series data management, due to the low time resolution of HSR sky survey.

\textbf{Distributed data store}. The in-memory stores had high throughput and low latency, and several solutions were considered before we embarked on AstroCache development. In-memory distribution file systems, such as Alluxio\cite{li2014tachyon}, was similar to HDFS\cite{HDFS} on disk. It followed the write-once, read-many approach for its files and applications, but sometimes we had to update data. In-memory distribution messaging systems, such as Apache Kafka\cite{ApacheKafka}, did not support random read. In-memory relational databases, such as MonetDB\cite{idreos2012monetdb}, have been tested for the insert performance through GWAC's catalogs\cite{wan2016column}. The experiment showed the insert latency was not stable, and sometimes an insert operation could not be finished within 15 seconds, because of periodically pushing data onto disk. In addition, it did not support query for scientific phenomenon data. In-memory distributed key-value stores, such as Redis cluster\cite{RedisCluster}, have been tested by us through GWAC's catalogs without our optimization. However, the network latency was the main bottleneck because of too many keys. The total amount of memory, which was consumed to keep one-night GWAC's catalogs into Redis cluster, was also unacceptable (more than data size), even if data was compressed.  The on-disk stores are suitable for long-term historical data storage. However, some popular solutions are not applicable to our scenario. Apache Cassandra\cite{Cassandra} is a distributed key-value database, but it is suitable for large-scale data storage. HDFS as a distributed file system, can support large-scale data storage but it does not support the necessary index and data organization schema. Hbase without our optimization is also tested by us through GWAC's catalogs. The insert latency is unacceptable in our cluster. In addition, Hbase does not support the scientific analysis methods.

\section{Summary}\label{section:Summary}
AstroServ is a distributed database to analyze and manage large-scale full life-cycle astronomical data for short-timescale and large field of view sky survey. Detect service on streaming can finish the detect of 3.5 million stars within 3.97 seconds. Real-time data service can finish the ingest of 3.5 million rows of data within 2.35 seconds. Long-term data service can finish 6.7 billion rows of data within 3.5 hours. The query latency on AstroCache is within 2.72 seconds and it on AstroBase is within 26 seconds. The overall performance is improved under our optimization. In future, our query will use polystore framework to automatically adjust data distribution for better performance.
\section{Acknowledgement}
This research was partially supported by the grants from the National Key Research and Development Program of China (No. 2016YFB1000602, 2016YFB1000603); the Natural Science Foundation of China (No. 91646203, 61532016, 61532010, 61379050, 61762082);  the Fundamental Research Funds for the Central Universities, the Research Funds of Renmin University (No. 11XNL010); and the Science and Technology Opening up Cooperation project of Henan Province (172106000077).
\bibliographystyle{plain}
\bibliography{astroDB}
\end{document}